# High-performance dendritic metamaterial absorber for broadband and near-meter wave radar

Song Jiaoyan[1], Zhao Jing[2]*, Li Yimin[1], Li Bo[1] and Zhao Xiaopeng[1]*

[1]*Smart Materials Laboratory, Department of Applied Physics, Northwestern Polytechnical University, Xi'an 710129, PR China*，[2]*Medtronic plc, Boulder, CO 80301, USA*

**Abstract**: Absorbing materials in ultra-high frequency (UHF) band has constantly been a major challenge. The size of the absorber in UHF band is large, whereas the resonant frequency band is narrow. According to Rozanov's theory, two kinds of composite metamaterial absorbers are designed to realize the requirements of low-frequency broadband metamaterial microwave absorber: the magnetic- metamaterial composite absorber1 (MA1) and the dielectric- metamaterial composite absorber 2 (MA2). In the range of approximately 300-1000MHz, both absorbers achieve absorption of over 90% and feature good adaptability to the incident angle of the incident wave. The absorbers also present good absorption rate of over 80% in the range of $0° − 45°$. Processing samples of indium tin oxide (ITO) resistance film and polymethacrylimide (PMI) foam board feature simple preparation and low cost, and the most important thing is to consider the weight problem, which features certain advantages in terms of use.
*Index Terms*: UHF, metamaterial, low-frequency broadband, dendritic resistance film

## I. Introduction

The application of ultra-high frequency (UHF) electro-magnetic(EM) wave has been widely used in many fields [1–3] nowadays. The military field, utilizes radar technology in low-frequency band; the civil field, commonly uses UHF band in radio frequency identification (RFID) technology, EM compatibility (EMC), and 4G mobile communication technology [4–6]. Further, EM radiation surrounds in people's lives. Therefore, the development of low-frequency absorbers can be used for radar stealth on aircraft surfaces and to effectively improve the EMC performance of application systems and the EM environment of humans. Absorbers are a kind of energy conversion material, that dissipates the incident EM energy or make it disappear by EM interference; they can absorb the incident EM wave effectively without reflection. Since Landy first proposed a perfect metamaterial absorber in 2008 [7], metamaterials has become a hot research topic in the field of wave absorption. This material is an artificial composite material with periodic structure unit with subwavelength size. Metamaterials also possess abnormal physical properties, such as negative permeability and negative permittivity, which are not possessed by

E-mail: xpzhao@nwpu.edu.cn, zhaojing1120@gmail.com, Tel.: +86-29-88431662

*Corresponding author: J. Zhao, X. Zhao

natural materials [8–12]. The use of metamaterials as absorber breaks through the thickness of traditional material λ/4, reduces the overall thickness, and broadens the working frequency band of the absorber [13–15], providing a broad prospect for the development of absorbing materials. The research methods for EM metamaterial absorbers have been gradually matured, and their achievements include various frequency bands of EM waves, however most results are concentrated in the GHz to higher frequency bands [16–18]. As the wavelength of UHF band is long, the corresponding absorber is generally larger, and the absorption bandwidth at resonance point is relatively narrow, resulting difficulties in exploring UHF absorption materials. As a result, broadband microwave metamaterial absorbers of this band have rarely been reported. Owing to the current development and practical needs of metamaterial absorbers, this paper explores the possibility of realizing the meta-material low-frequency absorber and its realization method based on the basic requirements of designing the low-frequency metamaterial absorbers and considering the widened working frequency band as the center.

Widening the working frequency band of absorber and reducing the thickness of materials are challenges in the field of EM wave [19]. In 2000, Rozanov proved this point theoretically and proposed that working bandwidth $\Delta\lambda$ and thickness $d$ are two opposite parameters [20]. For example, for a single-layer Salisbury screen absorber, good performance can be achieved only in a specific frequency band [21]. However, the design goal of microwave absorber is to achieve minimum reflectivity in the widest frequency range with minimum thickness of materials. Absorption bandwidth is constantly related to EM parameters and thickness of materials. The relationship is that a wider working bandwidth $\Delta\lambda$ indicates greater thickness of the corresponding absorbing material or larger real part of the magnetic permeability. In 2010, Bao et al. [13] showed two structures, which are composed of dendritic cells with different geometric dimensions and absorption band widths of 1.93 at the microwave frequencies. Based on this model, Gu et al. [14] improved absorption band widths to 2.35 GHz. In 2014, Wang et al. [15] proposed a lightweight broadband absorber based on resistance. Absorption reaches over 80% in the frequency range of 8–1. 7 GHz, with a 72% relative bandwidth. And Yoo et al. proposed a flexible metamaterial absorber reaching good absorption in a narrow band around 400 MHz, with $1/62\lambda$ thickness of the material [22]. In 2017, Zuo et al. proposed a miniaturized metamaterial absorber, which achieve good absorption in the 860–960 MHz band, and are used in RFID systems [23]. The absorber features a narrow operating band of around 100 MHz. In 2017, J Mou et al. adopted a four-layer non-Foster circuit board and a magnetic substrate to achieve ultra-wideband absorption in 160–1000 MHz [19].The use of precision circuit boards and high-permeability substrates considerably improves absorption, substantially increasing the production cost and weight of absorbers. Therefore, to make absorber in UHF band have high-performance broadband absorption, at the same time, light-weight and low-cost is a difficult issue need to be solve.

According to Rozanov's theory, this paper proposes two kinds of low-frequency broadband composite metamaterial absorbers based on the resistance film with dendritic waveguide (RFDW) proposed by Wang Mei et al. [24, 25]. Two composite metamaterial absorbers combine metamaterial unit cells of different structures and sizes [26, 27]; these absorbers achieve an absorption of over 90% in the range of approximately 300–1000 MHz. The first metamaterial absorber is the magnetic-metamaterial composite absorber 1 (MA1) which increases the magnetic permeability of materials as the microwave permeability of magnetic materials (MM) is no more than a single-digit order of magnitude. The second absorber is the dielectric-metamaterial composite absorber 2(MA2) which sacrifice the thickness, but whose weight is within an acceptable range by using a low-density dielectric material.

## II. Magnetic - metamaterial composite absorber

According to Rozanov's theory, the operating bandwidth Δλ of the absorber is related to thickness d and permeability μ :

$$|\ln\rho_0|(\lambda_{max} - \lambda_{min}) < 2\pi^2 \sum_i \mu_{s,i}\, d_i \qquad (1)$$

where $\rho_0$ refers to the voltage reflection coefficient; $\lambda_{max}$ and $\lambda_{min}$ denote the maximum and minimum operating wavelength respectively. $d_i$ and $\mu_{(s,\,i)}$ are the thickness and static permeability of the i-th layer of the multilayer slab respectively. Two kinds of low-frequency broadband absorbers are designed to adopt the sandwich structure in this paper. The first is MA 1, whose design is shown in Fig. 1. The first layer represents the impedance surface of the dendritic slot, which is employed to guide the current flow. The second layer is a MM substrate, which mainly produces magnetic dielectric loss. The third layer is a metal grounding plate (GP).

Fig. 1 shows the unit cell of MA1. The top layer is an indium tin oxide(ITO) metasurface engraved with two different sizes of dendritic structure. The yellow area is a film with a square resistance value, which is recorded as *ohmic*. As presented in Fig. 1b, the first branch length is a=0.18 * Li, the second branch length is b= 0.26 * Li, and the branch width is w=0.08 * Pi. The period of the unit structure is L, where L1 corresponds to the center of the period of the large branch, L2 refers to the period of the surrounding four small branches, and 2*x denotes the distance between metasurfaces of adjacent periods. A substrate in the middle with dielectric constant ε , magnetic permeability μ, and thickness *h* is used.

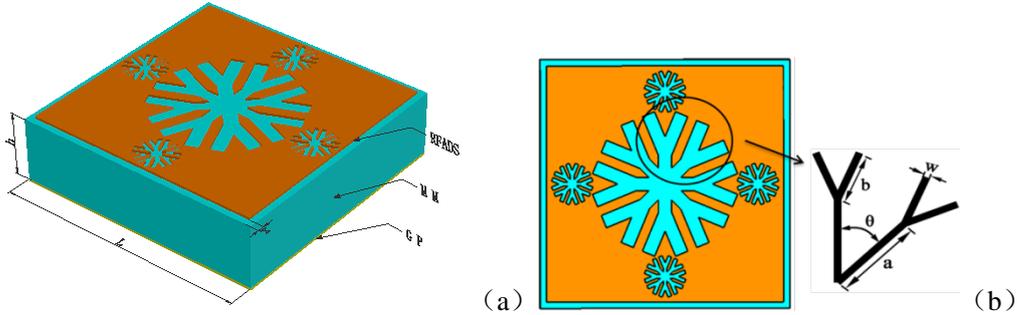

(a) (b)

**Fig. 1** Structure schematic of MA1 absorber **a** a single
unit cell **b** dendritic metasurface Structure

All simulations in the work are carried out using the commercial software Computer Simulation Technology (CST) Microwave Studio. Periodic boundary condition in the calculation process is used to simulate the actual materials (Fig. 2). The waveguide port with plane EM wave excitation is set, and the EM wave is travels the negative z axis through the model to the metal backboard. The frequency solver is used to calculate the reflection parameter ($S_{11}$). Transmission parameter is zero given the absence of transmission. Thus, we can obtain the absorption rate A(ω) from the equation: $A(\omega) = 1 - R(\omega) = 1 - |S_{11}|^2$.

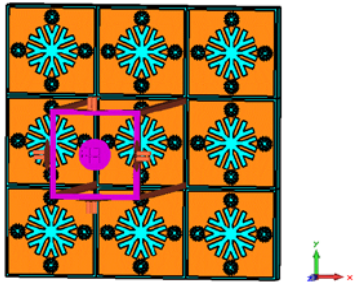

**Fig. 2** Periodic boundary setup of the model

A. Design, results, and discussion

For the MA1 model, design parameters are obtained through parameter optimization using CST and analyzing the influence of each geometric parameter on absorption. The best parameters include the following: L=105 mm, L1=70 mm, L2=20 mm, *ohmic* = 38 Ω/square, x = 2.7 mm, and θ=45°. The parameters of the middle magnetic substrate are ε=2, μ=3, h=27 mm. Copper is used as the metallic layer with a conductivity of 5.8×10^7 S/m. The thickness of copper is 0.036 mm. Fig. 3 shows the absorption curve of the model optimized by parameters. In the range of 0.32–1.1 GHz, the reflection coefficient consistently remains below −10 dB, that is, over 90% absorption. The nature of perfect absorption can be explained by the impedance matching theory[28]. At the absorption frequency, effective impedance is perfectly matched with free space( $Z = \sqrt{\mu(\omega)/\varepsilon(\omega)} = 1$ ) by controlling the electric permittivity ε and the magnetic permeability μ. Based on reflection and transmission parameters, effective impedance can be given by Ref.29:

$$Z = \sqrt{\frac{(1 + S_{11}(\omega))^2 - S_{21}^2(\omega)}{(1 - S_{11}(\omega))^2 - S_{21}^2(\omega)}} \qquad (2)$$

Fig. 3b plots the extracted real and imaginary parts of relative impedance. The real part of the relative impedance is closer to 1.0 at the absorption frequency.

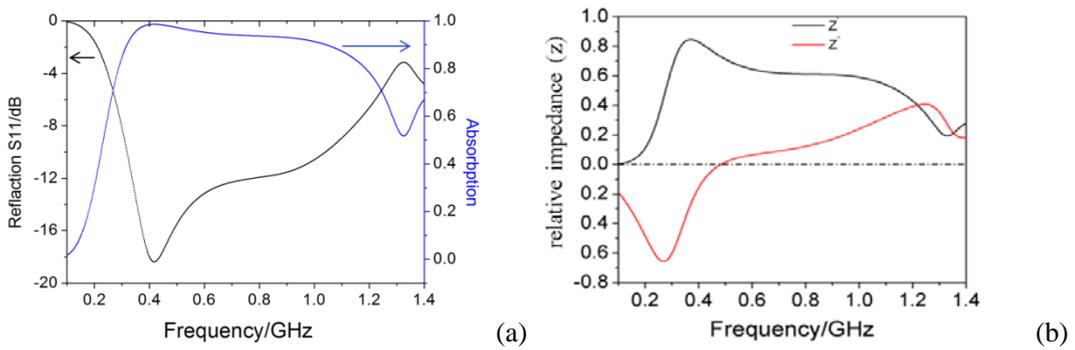

**Fig. 3** Simulation results of optimal structure MA1. **a** reflection characteristic curve S11; **b** relative wave impedance

To examine the importance of the dendritic metasurface and magnetic medium to MA1, we calculate the reflection characteristic curve S11 of the model of removed dendritic metasurface (Fig.4) and the model of removed magnetic medium (Fig.5) based on the optimal model. As shown in Fig. 4, when the dendritic metasurface is removed, the absorber of the single magnetic medium shows almost no absorption, indicateing the major role played by the dendritic metasurface. The below surface current map shows that the metasurface induces a current and produces electrical resonance, which produces magnetic resonance with the back metallic layers; the main loss in the absorber occurs on the dendritic metasurface. As shown in Fig. 5, when magnetic medium is changed to air ($\varepsilon=1$, $\mu=1$; blue area) medium, the absorption curve undergoes a blueshift, and the absorber still exhibits good absorption in the higher frequency part. The use of magnetic medium effectively reduces the operating frequency of the absorber.

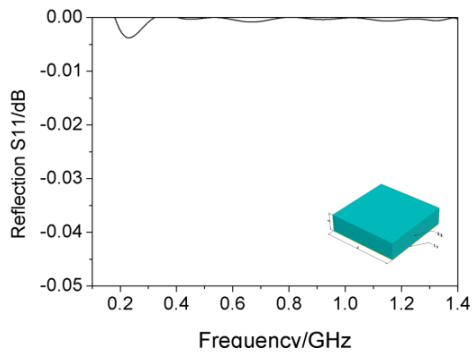 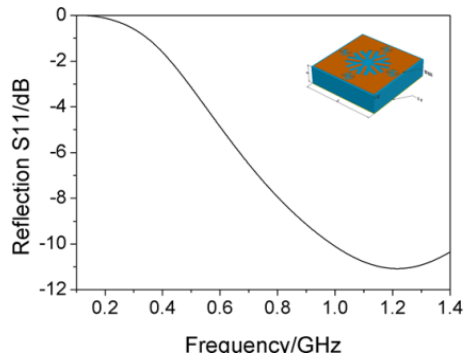

**Fig. 4** Reflection characteristic curve S11 when dendritic metasurface is removed

**Fig. 5** Reflection characteristic curve S11 when magnetic medium is removed

The dendritic metasurface uses semiconducting ITO film with the area resistance *ohmic*. Fig. 6 shows the reflection characteristic curve S11 for different values of area resistance *ohmic*. As *ohmic* increases from 18 Ω/square to 58 Ω/square, two absorption peaks combine as one peak, and a wide absorption band with reflectivity lower than −10 dB is obtained when *ohmic*=38 Ω/square, as shown by the blue line in Fig. 6.

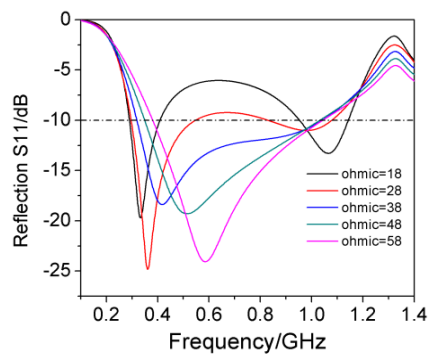

**Fig.6** Reflection characteristic curve S11 with different area resistances

The magnetic medium is a key point in the design of the absorber. Fig. 7 shows the influence of permeability μ on absorption. Different permeabilities μ correspond to different reflection characteristic curves (S11). As the value of permeability μ of the middle medium layer increases, the operating frequency of MA1 gradually shifts to the low-frequency region. Application of magnetic medium to the metamaterial absorber can effectively reduce the absorption frequency [30,31]. Given the dispersion characteristics of the actual magnetic material in the high-frequency part and the influence of magnetic permeability on weight, the optimum value μ=3 is considered.

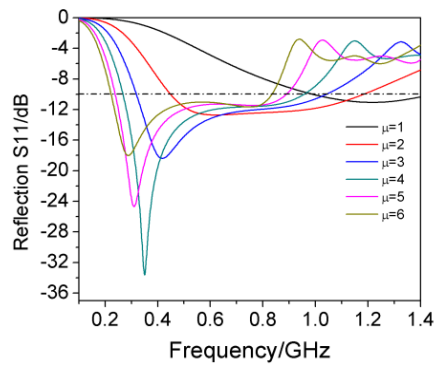

**Fig.7** Reflection characteristic curve S11 with different permeabilities

To further understand the absorption mechanism of MA1, we add surface current and energy flow monitors during simulation. Fig. 8a shows the surface current distribution of MA1 at 0.41 GHz. The red arrow indicates the direction of main current flow. The dendritic slot can guide the direction of surface current, and two circular electric currents with opposite directions are generated in the top layer benefit from this design. The two circular current flows can be equivalent to two dipoles, which generate considerable electric resonance. Moreover, the current in the back plate follows an opposite direction compared with that in the top layer, and this condition implies that a circular current can be formed between the top layer and back plate, which can yield strong magnetic resonance. As a result, EM resonance leads to EM energy absorption. The power loss distribution of MA1 at 0.4 GHz is also monitored, and the result are shown in Fig. 8b. Power loss is mainly dissipated in the horizontal direction between two small branches and the adjacent part to large branches.

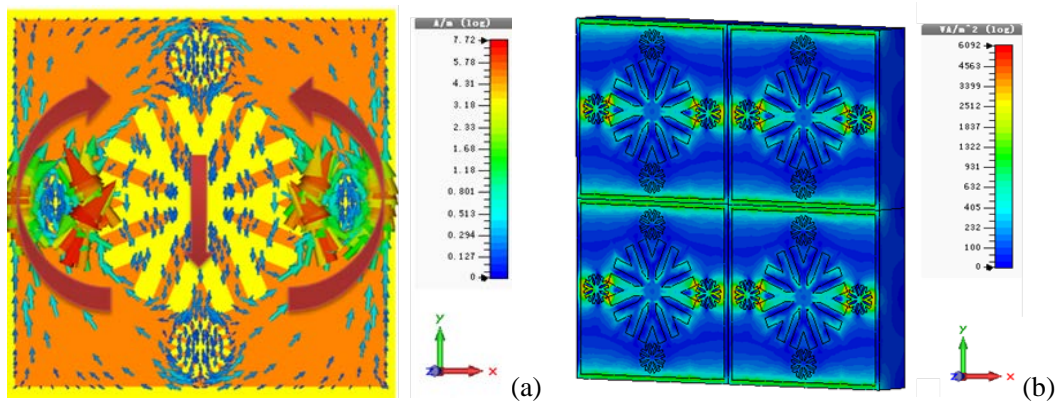

(a)      (b)

**Fig.8** The distribution of (**a**) surface current distribution and (**b**) power loss of MA1 at 0.4GHz

Owing to the high symmetry of MA1 structure, we understand that the MA1 model features polarization insensitivity characteristics. Now we analyze the influence of the angle of inclination of the incident EM wave on reflection. In this paper, the sensitivity of the MA1 model under the incident angle of transverse electric (TE) and transverse magnetic (TM) waves is simulated, as shown in Fig. 9. For the TE wave, as the incident angle is widens, the absorption performance gradually worsens (Fig. 9a). The MA1 model shows its eminent absorption character (>80%) at an incident angle of less than 45°. For the TM wave, the absorption character exhibits a complicated pattern. As shown in Fig. 9b, when the incident angle is less than 45°, the absorption in the low-frequency band (0.3–1 GHz) increases with incident angle, whereas the opposite is observed in the high-frequency band (1–1.4 GHz). When the incident angle is greater than 45°, the absorption performance gradually worsens as the incident angle is large.

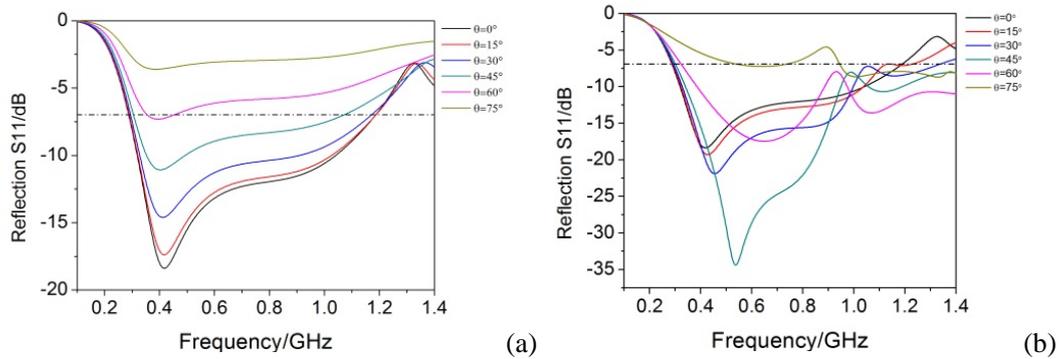

**Fig. 9** Reflection characteristic curve S11 at different incident angles for MA1. **a** TE wave **b** TM wave

### III. dielectric- metamaterial composite absorber

The second work is dielectric - metamaterial composite absorber 2 (MA2). The first layer is the impedance surface of the dendritic slot, which is similar to that of MA1. The second layer is polymethacrylimide (PMI) dielectric material, and the third layer is a metal GP. Fig. 10 shows the basic structure unit cell of MA2. The top layer is an ITO branch metasurface structure, and the dendritic structure adopts shade-engraved branches of two different periods. The whole unit cycle length is 2*L, and the width is L, whose period of large and small branch structure are L1 and L2, respectively. The middle is PMI dielectric substrate with a dielectric constant of 1.4 and possesses low density, and dielectric constant; its thickness is $h$. The bottom layer is copper with a thickness of 0.035 mm and a conductivity of 5.8 x $10^7$ S/m.

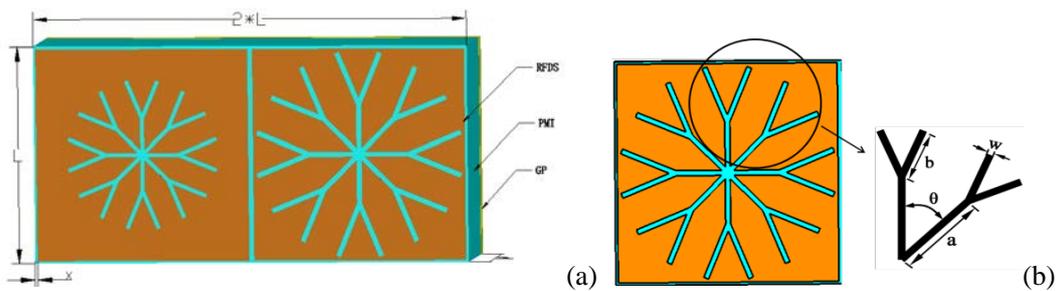

**Fig.10** Structure schematic of MA2 absorber **a** a single unit cell **b** dendritic metasurface structure

The EM environment set in the simulation calculation is the same as that of MA1, and the periodic boundary conditions are shown in Fig. 11.

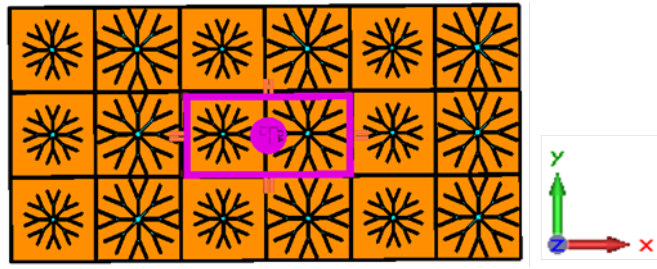

**Fig. 11** Periodic boundary setup of the model

B. Design, results, and Discuss

For the MA2 model, the design parameters are obtained through parameter optimization using CST and analyzing the influence of each geometric parameter on absorption. The best parameters of unit cell are as follows: L=110 mm, L1=110 mm, L2=80 mm, ε=1.4, μ=1, h=55 mm, *ohmic*=29 Ω/m2, w1=2.5 mm, w2=2 mm, and x=1.2 mm. Fig. 12a displays the absorption curve of the model optimized by parameters. In the frequency range of 0.38−1.05 GHz, the reflectivity of the absorber is less than −10dB, and the corresponding absorption rate exceeds 90%. Based on the reflection and the transmission parameters, effective impedance can be given, as shown in Fig. 12b.

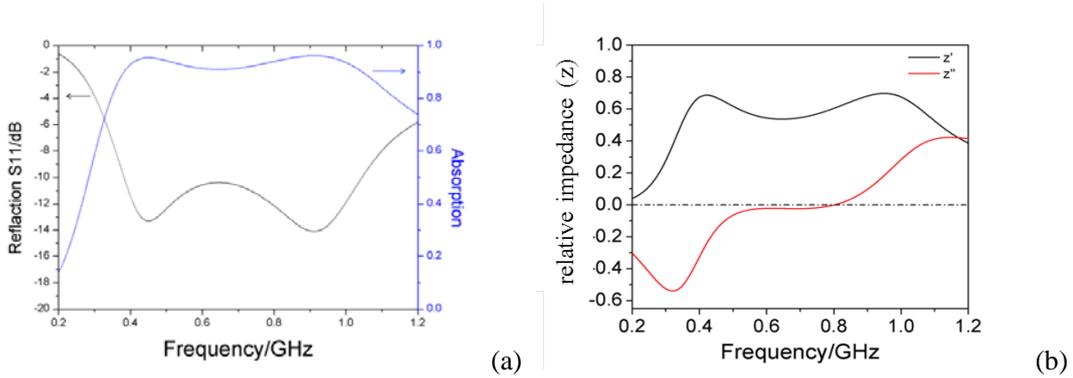

**Fig.12** Simulation results of optimal structure MA2. **a** reflection characteristic curve S11; **b** relative wave impedance

According to the combination of the models (Fig. 10), the unit cell of the absorber consists of two dendritic structures of different sizes, which absorb different microwave frequencies. Fig. 12a shows two absorption peaks at $f_1$=0.45 GHz and $f_2$=0.9 GHz. To explore the absorption mechanism, the structure of a single branch unit was simulated separately (Fig. 13). Fig.13a displays the reflection characteristic curve of the absorber consisting of a single small-branch structure unit. The central frequency band of the working response of the structure is f=0.38 Hz. Fig. 13b presents the reflection characteristic curve of the absorber consisting of a single large-branch structure unit. The central frequency band of the working response of the structure is f=0.71 GHz. Compared with the absorption characteristic curve of the composite structure, the absorption of the small branch unit corresponds to the first absorption peak, and that of the large branch unit corresponds to the second absorption peak. Moreover, after the combination of the two structural units, the working frequency of the center undergoes a blueshift, but the offset is negligible. As a result, the two absorption bands are connected, and the absorption bandwidth is widened. By combining different physical size metamaterial elements, the dendritic metamaterial units working in different frequency bands are combined to achieve continuous absorption in a wider frequency range.

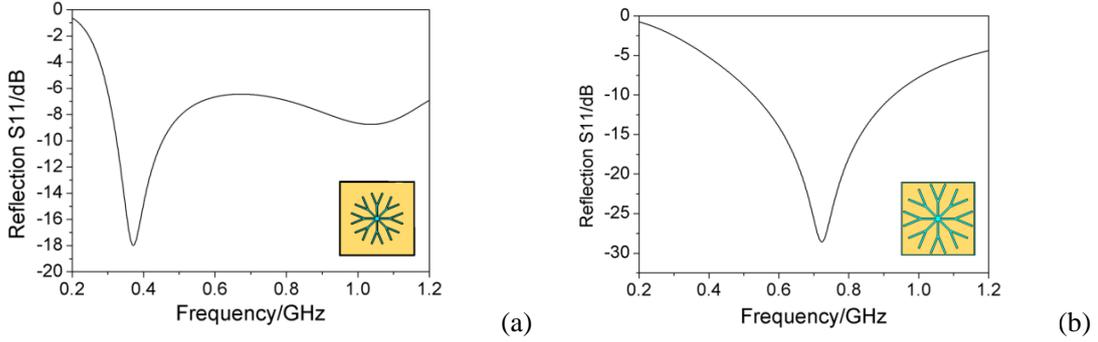

**Fig.13** Reflection characteristic curve S11 for **a** the absorber of single small unit cell and **b** the absorber of single large unit cell

To investigate the influence of surface resistance *ohmic* on the absorber, the simulated reflectivity for different values of *ohmic* is plotted in Fig. 14. As resistance *ohmic* increases from 21 Ω/square to 37 Ω/square, the reflective characteristic curve S11 changes from two reflective peaks to one, thus widening the absorption bandwidth. According to the principle of equivalent EM resonance, the quality factor Q of the equivalent circuit of the absorber is calculated as follows:

$$Q = \omega \frac{L}{R} = \frac{1}{\omega CR} \qquad (3)$$

where ω, L, C, and R denote the resonant angular frequency, equivalent inductance, equivalent capacitance and equivalent resistance respectively. When resonance occurs, the pass band BW is computed as follows :

$$BW = \frac{\omega}{Q} \qquad (4)$$

Formulas (3) and (4) show that when EM resonance occurs, the value of the quality factor Q decreases with increasing R value, and the sharp curve becomes smooth, such that the passband widens. When *ohmic*= 29 Ω/square, the two absorption peaks are connected together to obtain good absorption in a wider band and achieve the desired results ( i.e. the blue curve in the figure).

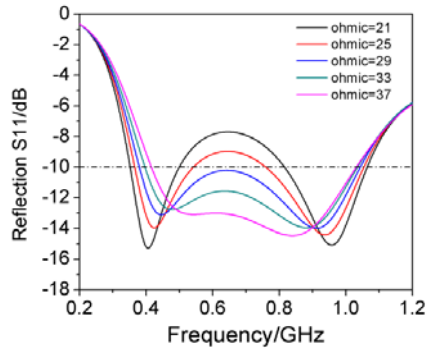

**Fig.14** Reflection characteristic curve S11 with different area resistances

To further understand the absorption mechanism of MA2, we add surface current and energy flow monitors during simulation. Fig.15a shows the surface current distribution of MA1 at 0.9 GHz. The red arrow indicates the direction of main current flow. Each dendritic unit forms a separate current loop. The current is dense near the branches，and the dendritic waveguides exert an induction effect. The distribution of energy flow in Fig. 15b shows that energy loss is concentrated near the branches, indicating that the dendritic structure of metamaterials plays an important role in the absorber.

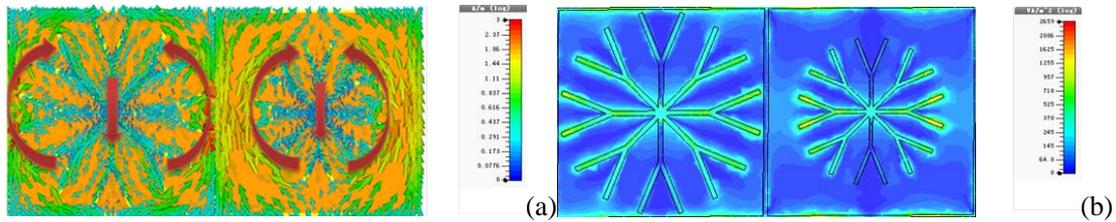

**Fig.15** The distribution of **a** surface current distribution and **b** power loss of MA2 at 0.9 GHz

Finally, considerable attention is focused on demonstrating the insensitivity of the absorber to the incident angle. Reflectivity curves under TE and TM incident waves with different incident angles for the MA2 are simulated and shown (Fig. 16). For TE wave, when the incident angle is less than 45°, the absorber still presents a good absorption effect and the absorption coefficient is over 80%.However, as the angle continually increases, the absorption effect worsens. For TM wave, the variation in absorption effect is more complicated. At an incident angle of less than 45°, as the incident angle increases, the absorption effect improves but slightly undergoes a blueshift. Over 45° range , poorer absorption effect is observed with increasing incident angle.

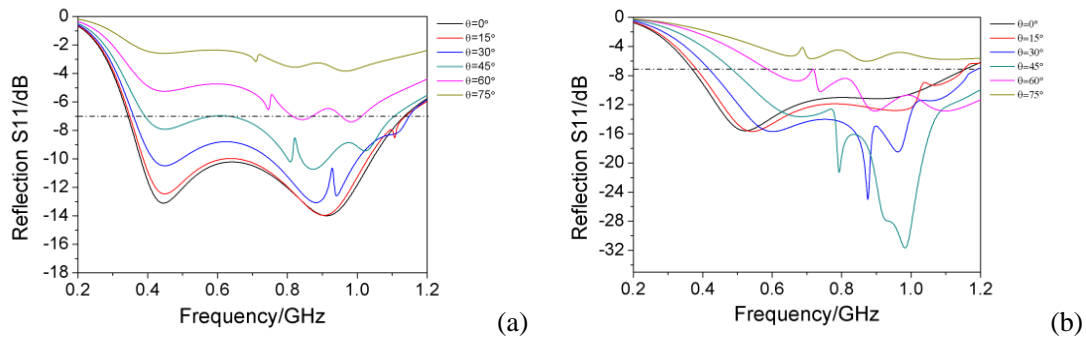

**Fig. 16** Reflection characteristic curve S11 at different incident angles for MA2.
**a** TE wave **b** TM wave

## IV. Conclusion

In this paper, two kinds of UHF band low-frequency broadband metamaterial absorbers are designed. Absorption rates exceed 90% in the range of approximately 300−1000 MHz. For MA1, magnetic medium is used to reduce the thickness of the middle substrate, and the absorption frequency band is 320−1100 MHz. For MA2, the thickness of the absorber is sacrificed, but the dielectric substrate PMI foam of low density is used; the absorption frequency band is 380−1050 MHz. Both ofabsorbers present good adaptability to the incident angle of the incident wave. At an incident angle of 45 °, both absorbers show good absorption rate of over 80 %，and are insensitive to the planned mode of incident wave. Compared with other absorbers, the propose absorbers feature a wide working frequency range of approximately 700 MHz. Processing samples of ITO resistance film and PMI foam board are simple and inexpensive. In designing absorbers, the most important thing is to consider the weight problem, which feature certain advantages in terms of use.

## ACKNOWLEDGMENTS

This work was supported by the National Natural Science Foundation of China (Grant Nos. 11674267) and the National Key Scientific Program of China (under project No. 2012CB921503).

**References**

[1] R. Abhari, G.V. Eleftheriades (2003) Metallo-dielectric electromagnetic bandgap structures for suppression and isolation of the parallel-plate noise in high-speed circuits. IEEE Trans. Microw. Theory Tech. **51**(6):1629-1639.

[2] K. Ito, N. Haga, M. Takahashi, et al (2012) Evaluations of Body-Centric Wireless Communication Channels in a Range From 3 MHz to 3 GHz. Proc. IEEE **100**(7):2356-2363.

[3] Y. Okano, S. Ogino, K. Ishikawa (2012) Development of Optically Transparent Ultrathin Microwave Absorber for Ultrahigh-Frequency RF Identification System. IEEE Trans. Microw. Theory Tech. **60**(8):2456-2464.

[4] N.Z. Chen, X. Qing, L.H. Chung (2009) A Universal UHF RFID Reader Antenna. IEEE Trans. Microw. Theory Tech. **57**(5):1275-1282.

[5] Kawano Y , Hayashida S , Bae S , et al (2005) A study on miniaturization of 900 MHz and 2 GHz band antennas utilizing magnetic material. Antennas & Propagation Society International Symposium IEEE **3B**:347-350.

[6] S. Gu, J.P. Barrett, T.H. Hand, et al (2010) A broadband low-reflection metamaterial absorber. J. Appl. Phys. **108**(6):064913.

[7] N.I. Landy, S. Sajuyigbe, J.J. Mock, D.R. Smith, W.J. Padilla (2008) Perfect metamaterial absorber. Phys. Rev. Lett. **100**(4):207402.

[8] J.B. Pendry, Negative refraction makes a perfect lens. Phy. Rev.Lett. **85**, 4 (2000)

[9] B.Q. Liu, X.P. Zhao, W.R. Zhu, W. Luo, X.C. Cheng (2008) Multiple pass-band optical left-handed metamaterials based on random dendritic cells. Adv. Funct. Mater. **18**(21):3523-3528.

[10] X.P. Zhao (2012) Bottom-up fabrication methods of optical etamaterials. J. Mater. Chem. **22**(19):9439-9449.

[11] Q. Zhao, X.P. Zhao, L. Kang (2004) The defect effect in the one-dimensional negative permeability material. Acta Phys. Sin. **53**(7):2206-2211.

[12] C.R. Luo, L. Kang, Q. Zhao (2005) Effect of nonuniform-defect split ring resonators, on the left-handed metamater ials. Acta Phys. Sin. **54**(4):1607-1612.

[13] S. Bao, C.R. Luo, Y.P. Zhang, X.P. Zhao, Broadband metamaterial absorber based on dendritic structure. Acta Phys. Sin. **59**, 5(2010)

[14] S. Gu, B. Su, X.P. Zhao (2013) Planar isotropic broadband metamaterial absorber. J. Appl. Phys. **114**(16):163702

[15] B. Wang, B.Y. Gong, M. Wang, B. Weng, X.P. Zhao (2014) Dendritic wideband metamaterial absorber based on resistance film. App. Phys. A **145**(5):1559-1563.

[16] W. R. Zhu, X. P. Zhao, B. Y. Gong, L. H. Liu, and B. Su (2011) Optical metamaterial

absorber based on leaf-shaped cells. Appl. Phys. A **102**(147):147-151.

[17] B.X. Khuyen, B.S. Tung, Y.J. Yoo, et al (2017) Miniaturization for ultrathin metamaterial perfect absorber in the VHF band. Sci. Rep. **7**:45151.

[18] B.Y. Gong, F. Guo, W.K. Zou, et al (2017) New design of multi-band negative-index metamaterial and absorber at visible frequencies. Mod. Phys. Lett. B **31**(11):1750286.

[19] J. Mou, Z. Shen (2017) Broadband and thin magnetic absorber with non-Foster metasurface for admittance matching. Sci. Rep. **7**(1):6922.

[20] K.N. Rozanov (2000) Ultimate thickness to bandwidth ratio of radar absorbers. IEEE Trans. Antennas Propag. **48**(8):1230-1234.

[21] W.W. Salisbury (1952) Absorbent body for electromagnetic waves..

[22] Y.J. Yoo, H.Y. Zheng, Y.J. Kim, et al (2014) Flexible and elastic metamaterial absorber for low frequency, based on small-size unit cell. Appl. Phys. Lett. **105**(4):041902-041902-4.

[23] W. Zuo, Y. Yang, X. He, et al (2017) A Miniaturized Metamaterial Absorber for Ultrahigh-frequency RFID System. IEEE Antennas Wireless Propagat. Lett. **16**:329-332.

[24] M. Wang, B. Weng, J. Zhao, et al (2017) Dendritic-metasurface-based flexible broadband microwave absorbers. App. Phys. A **123**(6):434.

[25] Y.H. Liu, X.P. Zhao (2014) Perfect Absorber Metamaterial for Designing Low-RCS Patch Antenna. IEEE Antennas and Wireless Propagat. Lett. **13**:1473-1476.

[26] W.R. Zhu, X.P. Zhao (2009) Metamaterial absorber with dendritic cells at infrared frequencies. J. Opt. Soc. Am. B **26**(26):2382-2385.

[27] Y.H. Liu, S.L. Fang, et al (2013) Multiband and broadband metamterial absorbers. Acta Phys. Sin. **62**(13):134102.

[28] T. Koschny, E.N. Economou, C. M. Soukoulis, et al (2004) Effective medium theory of left-handed materials. Phys. Rev. Lett. **93**(10):107402-0.

[29] D.R. Smith, D.C. Vier, T. Koschny, et al (2005) Electromagnetic parameter retrieval from inhomogeneous metamaterials. Phys. Rev. E Stat. Nonlin. Soft Matter Phys. **71**(3):36617-0.

[30] Q. Chen, S. Bie, W. Yuan, et al (2016) Low frequency absorption properties of a thin metamaterial absorber with cross-array on the surface of a magnetic substrate. J. Phys. D Appl. Phys. **49**(42):425102.

[31] B.H. Zhang, W.L. Deng, H.P. Zhou, et al (2013) Low frequency needlepoint-shape metamaterial absorber based on magnetic medium. J. Appl. Phys. **113**(1):013903.